\documentclass[11pt]{article}
\usepackage{mypre}
\usepackage{amssymb}
\usepackage[dvips]{graphicx}

\newcommand{\case}[2]{{\ensuremath{\textstyle\frac{#1}{#2}}}}

\begin{document}

\title{
	Remark on the Theoretical Uncertainty in $B^0$-$\bar{B}^0$ Mixing
}
\author{Andreas S. Kronfeld \\
{\em\small Theoretical Physics Department,
Fermi National Accelerator Laboratory, 
Batavia, Illinois 60510}
\and
Sin\'ead M. Ryan \\
{\em\small School of Mathematics, Trinity College, Dublin~2, Ireland}}
\date{June 5, 2002}	

\preprint{FERMILAB-Pub-02/109-T \\ hep-ph/0206058}

\maketitle
\begin{abstract}
	We re-examine the theoretical uncertainty in the Standard Model 
	expression for $B^0$-$\bar{B}^0$ mixing.
	We focus on lattice calculations of the ratio~$\xi$, needed
	to relate the oscillation frequency of $B^0_s$-$\bar{B}^0_s$
	mixing to the poorly known CKM element~$V_{td}$.
	We replace the usual linear chiral extrapolation with one that
	includes the logarithm that appears in chiral perturbation
	theory.
	We find a significant shift in the ratio~$\xi$, from the
	conventional $1.15\pm0.05$ to $\xi=1.32\pm0.10$.
\end{abstract}

It is anticipated that the oscillation frequency of 
$B^0_s$-$\bar{B}^0_s$ mixing will be measured during Run~2 of the 
Tevatron~\cite{Anikeev:2002bb}.
It is thus timely to assess the measurement's impact on tests of the
Cabibbo-Kobayashi-Maskawa (CKM) picture of flavor and $CP$ violation.
The CKM interpretation is limited by the poorly known hadronic
matrix elements for $B^0_s\leftrightarrow\bar{B}^0_s$ and
$B^0_d\leftrightarrow\bar{B}^0_d$ transitions.
In this paper we re-examine lattice calculations of these matrix
elements, focusing on the chiral extrapolation.
We find that the range usually quoted is probably incorrect.

In the Standard Model, the theoretical expression for the oscillation
frequency is
\begin{equation}
	\Delta m_q = \left(\frac{G_F^2m_W^2S_0}{16\pi^2m_{B_q}}\right) 
		|V_{tq}^*V_{tb}|^2 \eta_B {\cal M}_q ,
	\label{eq:Delta}
\end{equation}
where $q\in\{d,\,s\}$, $S_0$ is an Inami-Lim function, 
$\eta_B$ is a short-distance QCD correction, and
${\cal M}_q$ is the hadronic matrix element for
$B^0_q\leftrightarrow\bar{B}^0_q$ transitions.
In Eq.~(\ref{eq:Delta}), the parentheses consists of 
accurately known quantities, and $|V_{tq}^*V_{tb}|$ is the CKM factor.
The hadronic matrix element 
\begin{equation}
	{\cal M}_q=\langle\bar{B}_q^0| [\bar{b}\gamma^\mu (1-\gamma^5)q]
		[\bar{b}\gamma_\mu (1-\gamma^5)q]|B_q^0\rangle
\end{equation}
and $\eta_B$ depend on the renormalization scheme, but the product
$\eta_B{\cal M}_q$ does not.
The renor\-malization-group invariant value of the short-distance factor
is $\hat{\eta}_B=0.55$.

One should keep in mind that non-Standard physics at short distances
can modify Eq.~(\ref{eq:Delta}).
For convenience we shall couch the discussion as using $\Delta m_q$
and the hadronic matrix element to determine~$|V_{tq}|$.
The resulting value of~$|V_{tq}|$ can then be compared to other CKM 
determinations to test for deviations from the Standard Model.

${\cal M}_q$ must be computed with a non-perturbative method,
such as lattice gauge theory.
For historical reasons one usually writes
\begin{equation}
	{\cal M}_q = \frac{8}{3}m^2_{B_q}f^2_{B_q} B_{B_q}
\end{equation}
and focuses on the decay constants $f_{B_q}$ and 
the bag parameters~$B_{B_q}$.
But lattice QCD gives ${\cal M}_q$ directly
(and $f_{B_q}$ separately from 
$\langle 0|\bar{b}\gamma_\mu\gamma^5 q|B_q^0\rangle$).
The separation does, however, turn out to be useful, as we shall see 
below, when considering the dependence of $f_{B_q}$ and $B_{B_q}$ on 
the masses of the light quarks.

At present the uncertainty in the matrix elements is large.
A~recent review~\cite{Ryan:2001ej} of lattice calculations quotes
\begin{eqnarray}
	f_{B_s} = 230 \pm 30~\textrm{MeV} , & \quad &
	\hat{B}_{B_s} = 1.34 \pm 0.10 , \label{eq:fBs} \\
	f_{B_d} = 198 \pm 30~\textrm{MeV} , & \quad &
	\hat{B}_{B_d} = 1.30 \pm 0.12 . \label{eq:fBd}
\end{eqnarray}
These estimates take into account the first (partially) unquenched 
calculations of $f_{B_q}$ \cite{Collins:1999ff,AliKhan:2000eg,%
AliKhan:2001jg,Yamada:2001xp,Bernard:2001wy}, 
several quenched calculations of $B_{B_q}$ and preliminary results 
suggesting that $B_{B_q}$ changes little when the quenched approximation 
is removed~\cite{Yamada:2001xp}.
The raw Monte Carlo data in lattice calculations are generated with
the light quark mass~$m_q$ in the range 0.2--$0.5<m_q/m_s<1$,
and the physical matrix elements are obtained by extrapolating $m_q$
to the down quark's mass~$m_d$.
This method of reaching physically light quarks is called the chiral 
extrapolation, and it plays an important role in our analysis below.

The frequency for $B^0_d$-$\bar{B}^0_d$ mixing has been measured
precisely, $\Delta m_d=0.494\pm0.007~\textrm{ps}^{-1}$~\cite{mixing}.
With Eqs.~(\ref{eq:Delta}) and~(\ref{eq:fBd}) the uncertainty 
on~$|V_{td}|$ is limited to~15\% by $f_{B_d}\sqrt{B_{B_d}}$.
The precision on $|V_{td}|$ will not improve until better (unquenched)
lattice calculations have been carried out.
The frequency for $B^0_s$-$\bar{B}^0_s$ mixing is known to be high, 
$\Delta m_s>15~\textrm{ps}^{-1}$~\cite{mixing}.
But details of the way $\Delta m_s$ is extracted from the data mean
that the first measurement will immediately have a precision at the
percent level~\cite{Anikeev:2002bb}.
Thus, it is interesting to form the ratio
\begin{equation}
	\frac{\Delta m_s}{\Delta m_d} = 
		\left|\frac{V_{ts}}{V_{td}}\right|^2
		\frac{m_{B_s}}{m_{B_d}} \xi^2 ,
	\label{eq:ratio}
\end{equation}
where
\begin{equation}
	\xi^2 = \frac{f_{B_s}^2 B_{B_s}}{f_{B_d}^2 B_{B_d}} ,
\end{equation}
and use Eq.~(\ref{eq:ratio}) to determine $|V_{td}|$.
The measurement uncertainties are (or soon will be) negligible.
By CKM unitarity $|V_{ts}|=|V_{cb}|$ to good approximation.
Thus, the error in $|V_{td}|$ is
\begin{equation}
	\delta|V_{td}| = \sqrt{ (\delta|V_{cb}|)^2 + (\delta\xi)^2 }.
\end{equation}
The uncertainty in $|V_{cb}|$, determined from semileptonic $B$ decay,
is also dominated by QCD, but it is only 2--4\% and relatively well 
understood~\cite{Cronin:2001fk,Hashimoto:2001nb,Briere:2002ew}.

The conventional wisdom, coming from several reviews
of lattice $B$ physics, is that $\delta\xi$ is small.
Based on such endorsement, recent efforts to fit a wide range of
precisely measured flavor observables have used 
$\xi=1.14 \pm 0.03 \pm 0.05$~\cite{Ciuchini:2000de} or 
$\xi=1.16 \pm 0.03 \pm 0.05$~\cite{Hoecker:2001xe}.
The second error bar is meant to reflect the uncertainty from the 
quenched approximation; the first covers all other sources of 
uncertainty in lattice calculations.
Central values in this range are reproduced by many quenched, and 
some unquenched, calculations.

Such a small error is, however, not universally accepted in the
lattice community.
Booth~\cite{Booth:1994hx}, noting that chiral logarithms in the 
quenched approximation differ strikingly from those of QCD, predicted 
that $\xi$ in QCD would be 0.15--0.28 larger than in the quenched 
approximation.
Sharpe and Zhang~\cite{Sharpe:1996qp}, with a similar point of view, 
reckoned that $\delta(\xi-1)/(\xi-1)$ could be~100\%.
Bernard, Blum and Soni~\cite{Bernard:1998dg} studied two different 
ways of carrying out the analysis.
Treating $f_{B_s}/f_{B_d}$ and $B_{B_s}/B_{B_d}$ separately (as usual), 
they found $\xi=1.17 \pm 0.02^{+0.12}_{-0.06}$;
treating instead ${\cal M}_s/{\cal M}_d$ directly, 
they found $\xi=1.30 \pm 0.04^{+0.21}_{-0.15}$.
(In Ref.~\cite{Bernard:1998dg} the second error comes from studying 
the lattice spacing dependence; the difference is significant
source of concern~\cite{Soni:2002rm}.)
Finally, the JLQCD collaboration studied the effect of the
chiral log in lattice calculations with two light flavors,
finding that the extrapolated value of~$\xi$ could change
significantly~\cite{Yamada:2001xp}.

At first glance, $\delta\xi/\xi$ could well be smaller than 
$\delta f_{B_q}/f_{B_q}$.
$\xi$~is a ratio of similar quantities, so, in numerical lattice
calculations, most of the Monte Carlo statistical fluctuations
do cancel.
Similarly, the short-distance normalization factor of the lattice 
operator also cancels.
But one is still left with a multi-scale problem, with
the heavy quark mass~$m_b$, 
the QCD scale~$\Lambda_{\mathrm{QCD}}$ 
and the range of light quark masses from~$m_s$ down to~$m_d$.
Because the numerator and denominator of $\xi$ are the same, except
for the light quark, one may expect $\xi$ to be insensitive to the
heavy-quark and QCD scales, but not to scales between $m_s$ and~$m_d$.

Let us examine the uncertainties associated with each scale in more
detail.
Heavy-quark corrections to $\xi$ are suppressed
by~$(m_s-m_d)/m_b\sim2\%$.
In lattice calculations, one should also worry about discretization 
effects of the heavy quark, because $m_ba\sim1$.
There are several ways to handle this problem and some debate over
the best method~\cite{Kronfeld:2000id}.
But the various discretizations yield consistent results 
for $f_{B_s}/f_{B_d}$ and~$B_{B_s}/B_{B_d}$.
Thus, we conclude that errors from the short distance scales are under 
control.

Next let us consider $\Lambda_{\mathrm{QCD}}$.
Implicit in the quenched approximation (also called the valence 
approximation) is that the omitted sea quarks are compensated by a 
shift in the bare gauge coupling~\cite{Weingarten:1981jy}.
This treats light-quark vacuum polarization in a dielectric
approximation.
Such approximations can be accurate when looking at a narrow range
of scales.
In the case at hand, that means that ratios of decay constants or bag
parameters could be accurate as long as all quark masses are not too
different.
Thus, it is plausible that the quenched approximation accurately
determines the slope of $\xi$, viewed as a function of $r=m_q/m_s$,
when $r\sim 1$.
Unquenched calculations \cite{Collins:1999ff,AliKhan:2000eg,%
AliKhan:2001jg,Yamada:2001xp,Bernard:2001wy}
do not contradict this expectation.
These calculations, and the justification of the quenched
approximation~\cite{Weingarten:1981jy}, suggest that the scale
$\Lambda_{\mathrm{QCD}}$ is also under control.

That leaves us with contributions to~$\xi$ from the long distances 
between $1/m_s$ and $1/m_d$.
Here the quenched approximation is known to break
down~\cite{Booth:1994hx,Sharpe:1996qp}, and it is not obvious that
the quenching error could be as small as~5\%.
One must take a careful look at how the chiral extrapolation is done,
and consider what methods of extrapolation are reliable.

The correct framework to discuss the long-distance behavior of QCD, and
the chiral extrapolation in particular, is chiral perturbation theory.
We neglect $1/m$ corrections and write
\begin{eqnarray}
	\sqrt{m_{B_q}} f_{B_q} & = & \Phi \left[1 + \Delta f_q \right],\\
	               B_{B_q} & = &   B  \left[1 + \Delta B_q \right],
\end{eqnarray}
where $\Phi$ and $B$ are independent of both heavy and light quark
masses, and $\Delta f_q$ and $\Delta B_q$ denote the (one-loop)
contribution of the light meson cloud.
The ``chiral logarithms'' reside in $\Delta f_q$ and~$\Delta B_q$.

Neglecting isospin breaking,
the one-loop corrections to the decay constants
are~\cite{Grinstein:1992qt,Goity:1992tp,Booth:1994hx,Sharpe:1996qp}
\begin{eqnarray}
	\Delta f_s = - \frac{1+3g^2}{(4\pi f)^2} \left[
					 m_K^2    \ln\left(m_K^2/\mu^2\right)
			\right.  & + & \left.
		\case{1}{3}  m_\eta^2 \ln\left(m_\eta^2/\mu^2\right) \right] 
		\nonumber \\ & + &
		\left(m_K^2 + \case{1}{2} m_\pi^2\right) f_1(\mu) +
		\left(m_K^2 - \case{1}{2} m_\pi^2\right) f_2(\mu) ,
	\label{eq:unquenchedfBs} \\
	\Delta f_d = - \frac{1+3g^2}{(4\pi f)^2} \left[
		\case{3}{4}  m_\pi^2  \ln\left(m_\pi^2/\mu^2\right)
			\right. & + & \left.
		\case{1}{2}  m_K^2    \ln\left(m_K^2/\mu^2\right) + 
		\case{1}{12} m_\eta^2 \ln\left(m_\eta^2/\mu^2\right) \right] 
		\nonumber \\ & + &
		\left(m_K^2 + \case{1}{2} m_\pi^2\right) f_1(\mu) +
		\case{1}{2} m_\pi^2 f_2(\mu) ,
	\label{eq:unquenchedfBd}
\end{eqnarray}
and to the bag parameters
\begin{eqnarray}
	\Delta B_s = - \frac{1-3g^2}{(4\pi f)^2} 
		\case{2}{3}  m_\eta^2 \ln\left(m_\eta^2/\mu^2\right) & + & 
		\left(m_K^2 + \case{1}{2} m_\pi^2\right) B_1(\mu) +
		\left(m_K^2 - \case{1}{2} m_\pi^2\right) B_2(\mu) ,
	\label{eq:unquenchedBBs} \\
	\Delta B_d = - \frac{1-3g^2}{(4\pi f)^2} \left[
		\case{1}{2}  m_\pi^2  \ln\left(m_\pi^2/\mu^2\right)
			\right. & + & \left.
		\case{1}{6}  m_\eta^2 \ln\left(m_\eta^2/\mu^2\right) \right] 
		\nonumber \\ & + &
		\left(m_K^2 + \case{1}{2} m_\pi^2\right) B_1(\mu) +
		\case{1}{2} m_\pi^2 B_2(\mu) ,
	\label{eq:unquenchedBBd}
\end{eqnarray}
where $f$ and $g$ are (the chiral limit of) the light pseudoscalar
decay constant and $B$-$B^*$-$\pi$ coupling.
The ``low-energy'' constants $f_i(\mu)$ and $B_i(\mu)$ encode QCD
dynamics from distances shorter than $\mu^{-1}$, whereas the logarithms
are long-distance properties of QCD, constrained by chiral symmetry.
The dependence on $\mu$ cancels in the total.

It is convenient to look separately at the $f_B$ and $\sqrt{B_B}$
factors in $\xi$.
The chiral logarithm in the $\sqrt{B_B}$ factor could be small because
it is multiplied by $1-3g^2$.
On the other hand, the chiral logarithm in the $f_B$ factor could be
significant, because it is multiplied by $1+3g^2$.
Consequently, we focus on
\begin{equation}
	\xi_f = f_{B_s}/f_{B_d}
	\label{eq:xif}
\end{equation}
and study its chiral extrapolation.
Our strategy is to use lattice calculations as an (indirect) way of 
determining the low-energy constants, and then we reconstitute~$\xi_f$.
Repeating our analysis for the chiral extrapolation of
$\xi_B=\sqrt{B_{B_s}/B_{B_d}}$ verifies that $\xi_B$ has a small effect.

\begin{figure}[p]
	\centering
 	\includegraphics[width=0.8\textwidth]{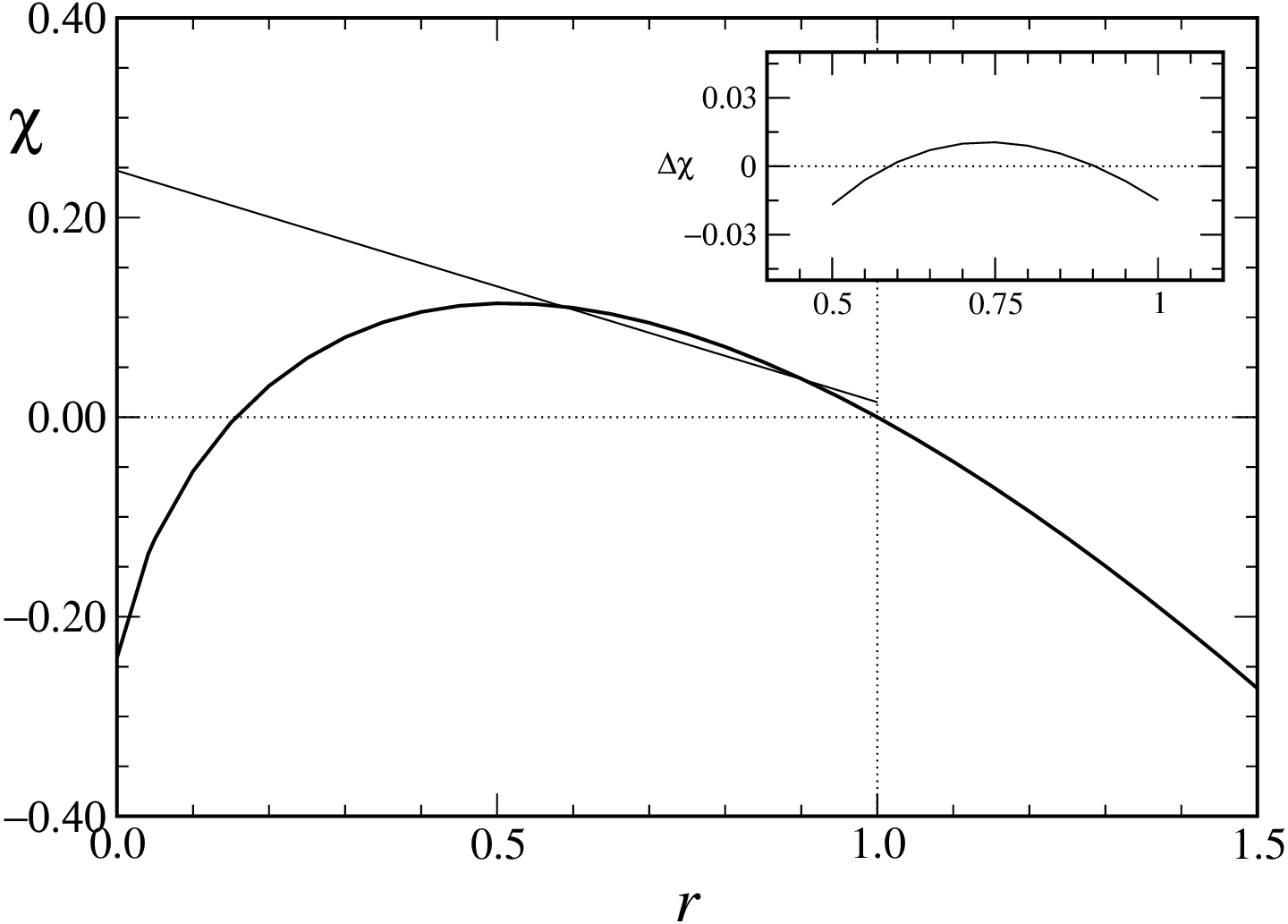}
	\caption[fig:chi]{Plot of the chiral logarithm $\chi(r)$
	as the mass ratio $r=m_{qq}^2/m_{ss}^2=m_q/m_s$ is varied,
	compared with a straight line fit for $0.5\le r\le 1.0$.
	The difference between the curve and the fit is shown in
	the inset.}
	\label{fig:chi}
\end{figure}

\begin{figure}[p]
	\centering
 	\includegraphics[width=0.8\textwidth]{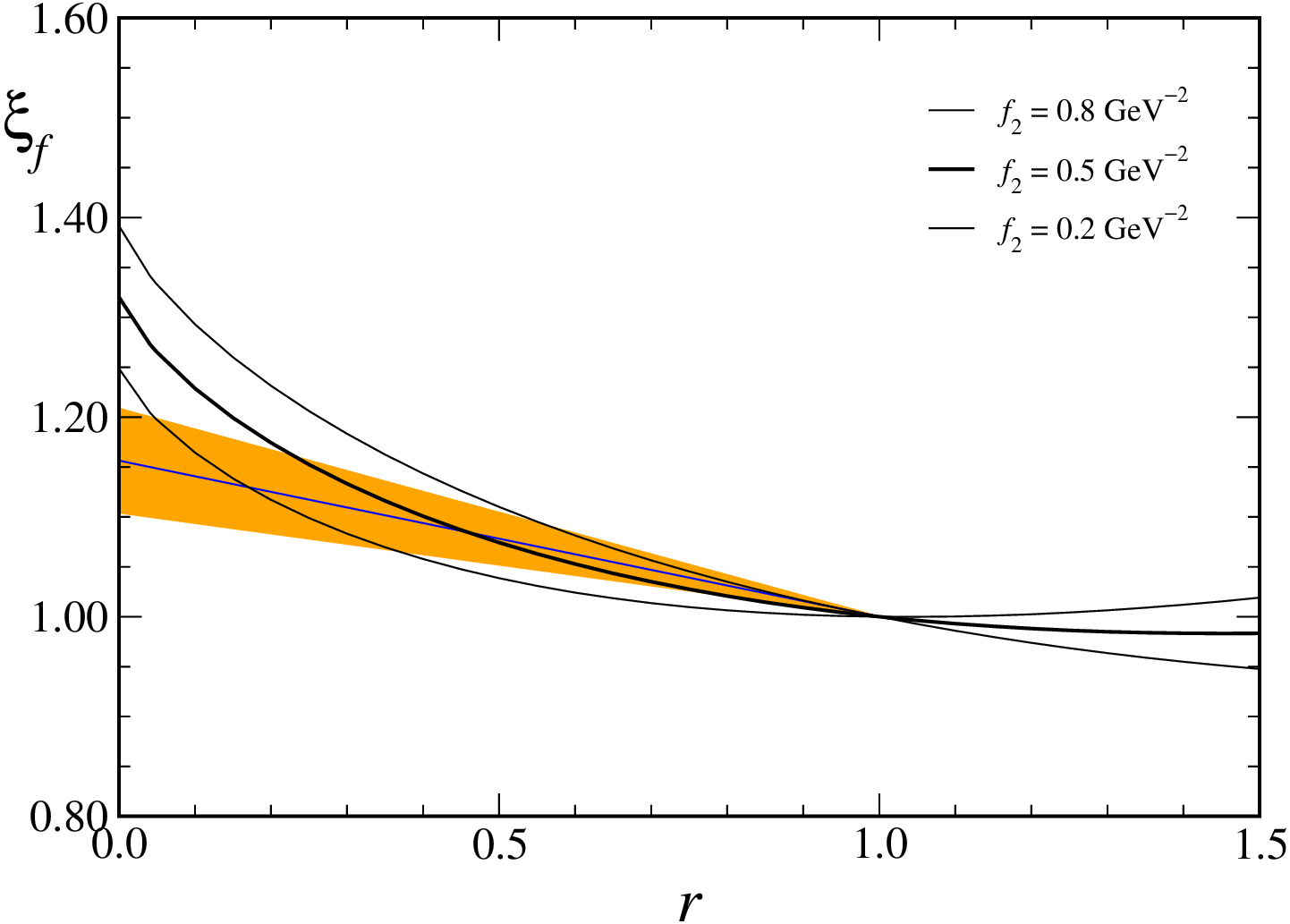}
	\caption[fig:xi_from_f2]{Plot of $\xi_f$ against $r$ for several
	values of the low-energy constant:
	$f_2(1~\textrm{GeV})=0.2,\,0.5,\,0.8~\textrm{GeV}^{-2}$.
	Also shown is the linear extrapolation with $\xi_f(r_d)=1.15\pm0.05$.}
	\label{fig:xi_from_f2}
\end{figure}

Combining Eqs.~(\ref{eq:unquenchedfBs}) and~(\ref{eq:unquenchedfBd}), 
the first non-trivial order in the chiral expansion is
\begin{equation}
	\xi_f - 1 = (m_K^2 - m_\pi^2) f_2(\mu) -
		\frac{1+3g^2}{(4\pi f)^2} \left[
		\case{1}{2}  m_K^2    \ln(m_K^2/\mu^2) + 
		\case{1}{4}  m_\eta^2 \ln(m_\eta^2/\mu^2) - 
		\case{3}{4}  m_\pi^2  \ln(m_\pi^2/\mu^2) \right] .
	\label{eq:xifchi}
\end{equation}
All lattice estimates of $\xi$ are obtained not at physical light 
meson masses, but by chiral extrapolation.
Therefore, we use Gell-Mann--Okubo formulae to replace the meson 
masses with
\begin{eqnarray}
	m_\pi^2  & = &   m_{qq}^2,               \label{eq:mpiGMO} \\
	m_K^2    & = &  (m_{ss}^2 + m_{qq}^2)/2, \label{eq:mKGMO}  \\
	m_\eta^2 & = & (2m_{ss}^2 + m_{qq}^2)/3. \label{eq:metaGMO}
\end{eqnarray}
Varying the light quark mass changes $m_{qq}^2\propto m_q$.
Lattice calculations typically take $m_{qq}^2$ not too different from
$m_{ss}^2$, so we write $m_{qq}^2=rm_{ss}^2$.
Then
\begin{equation}
	\xi_f(r) - 1 = m_{ss}^2(1 - r) \left\{\case{1}{2}f_2(\mu) -
		\frac{1+3g^2}{(4\pi f)^2} \left[\frac{5}{12}
			\ln(m_{ss}^2/\mu^2) + l(r) \right]\right\},
	\label{eq:xifr}
\end{equation}
where
\begin{equation}
	l(r) = \frac{1}{1-r}\left[
		\frac{1+r}{4}  \ln\left(\frac{1+r}{2}\right) + 
		\frac{2+r}{12} \ln\left(\frac{2+r}{3}\right) - 
		\frac{3r}{4}   \ln(r) \right].
	\label{eq:chilog}
\end{equation}
The function $\chi(r)=(1-r)l(r)$ contains the chiral logarithms.
It is plotted in Fig.~\ref{fig:chi}.
The curvature over $0.5\le r\le 1.0$ is too small to be resolved when
there are percent-level statistical uncertainties on $\xi_f$.
But once $r\ll1$, which is appropriate for the down quark with 
$r_d\approx1/25$, the curvature required by the chiral log has a
significant effect.
Fig.~\ref{fig:xi_from_f2} shows this effect,
comparing the conventional linear chiral extrapolation with
Eq.~(\ref{eq:xifr}), for~$f_2(\mu)$ in the range coming from
Eq.~(\ref{eq:f2S}), below.

When $\xi$ is calculated in lattice gauge theory, the range of $r$ 
is restricted to $r\lesssim1$ but $r\not\ll1$.
Usually, it is fit to a straight line
\begin{equation}
	\xi_f(r) - 1 = (1-r)S_f
	\label{eq:xifS}
\end{equation}
and similarly $\xi^2_B(r)-1=(1-r)S_B$.
Usually one assumes this linear extrapolation holds down to the chiral
limit, quoting $\xi=[1+(1-r_d)S_f][1+\case{1}{2}(1-r_d)S_B]$.
The chiral log says, however, that this procedure is not trustworthy.
It has been employed because there was, until recently, no independent
reliable estimate of the $B$-$B^*$-$\pi$ coupling~$g^2$ in the
coefficient of the chiral log.

The CLEO collaboration has recently measured the width of
the $D^*$ meson, which yields a value for the $D$-$D^*$-$\pi$
coupling~\cite{Anastassov:2001cw}.
Heavy-quark symmetry suggests that the $B$-$B^*$-$\pi$ coupling is
nearly the same.
On this basis, we shall set $g^2=0.35$, although below we allow for
20\% deviations.
With $g^2=0.35$, the chiral log in $\xi_B$ is truly small, because
$1-3g^2=-0.05$, but the chiral log in $\xi_f$ is multiplied with
$1+3g^2=+2.05$.

With this handle on $g^2$, we can interpret the lattice results
for~$S_f$ as a calculation of~$f_2(\mu)$.
We assume the linear fit given by Eq.~(\ref{eq:xifS}) makes sense
around $r=r_0\sim 1$, even though we do not trust it when $r\ll1$.
So, at $r_0$ we set the right-hand side of Eq.~(\ref{eq:xifr}) equal to
the right-hand side of Eq.~(\ref{eq:xifS}) and find
\begin{equation}
	m_{ss}^2 \case{1}{2} f_2(\mu) = S_f + m_{ss}^2
		\frac{1+3g^2}{(4\pi f)^2} \left[\frac{5}{12}
			\ln(m_{ss}^2/\mu^2) + l(r_0) \right].
	\label{eq:f2S}
\end{equation}
Then, inserting this result into Eq.~(\ref{eq:xifr})
\begin{equation}
	\xi_f(r) - 1 = (1 - r) \left\{S_f + m_{ss}^2 
	\frac{1+3g^2}{(4\pi f)^2} \left[ l(r_0) - l(r) \right] \right\}.
	\label{eq:xiffinal}
\end{equation}
To evaluate the right-hand side, one needs estimates of $f$,
$g^2$ and~$S_f$.
We use $f=130$~MeV and $g^2=0.35$.
In addition, we take~\cite{Ryan:2001ej}
\begin{equation}
	(1-r_d)S_f=0.15\pm0.05
	\label{eq:latinput}
\end{equation}
which brackets many quenched calculations (for which there is a lot
of experience and reproducibility) as well as less well-developed
unquenched calculations.%
\footnote{In fact, some ``unquenched'' calculations only have $n_f=2$.}

\begin{figure}[btp]
	\centering
 	\includegraphics[width=0.8\textwidth]{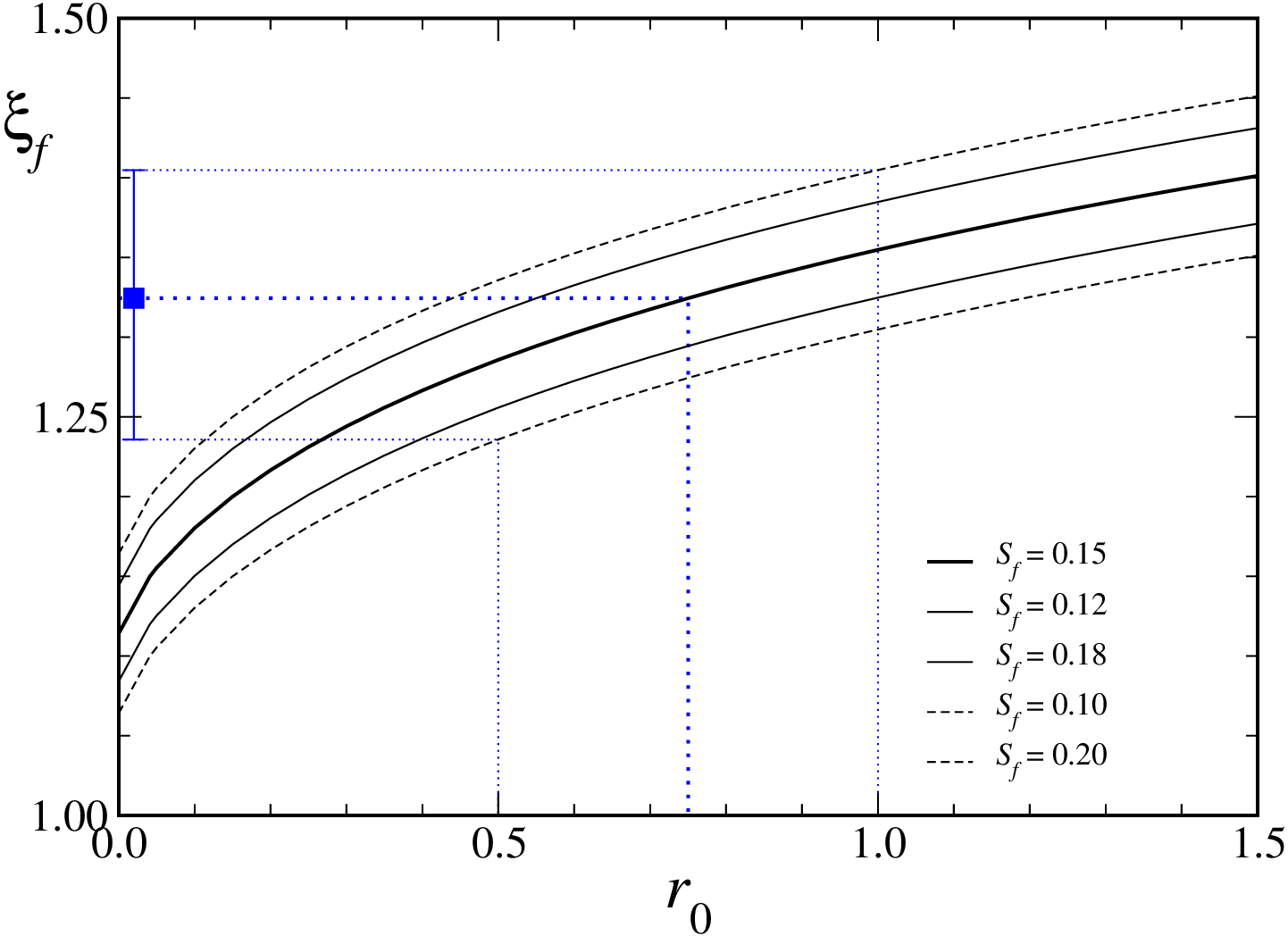}
	\caption[fig:xi]{Plot of $\xi_f$ from Eq.~(\ref{eq:xiffinal}),
	with $(1-r_d)S_f=0.15\pm0.05$, $m_{ss}^2=2(m_K^2-m_\pi^2)$,
	and $r=r_d=m_\pi^2/m_{ss}^2$, as a function of $r_0$.}
	\label{fig:xi}
\end{figure}

Once we have made the Ansatz to use the slope from lattice QCD to
determine the low-energy constant via Eq.~(\ref{eq:f2S}), another
source of uncertainty is the choice of~$r_0$.
Fig.~\ref{fig:xi} shows the result from Eq.~(\ref{eq:xiffinal})
for the physical value $\xi_f(r_d)$,
as a function of $r_0$ from 0~to~1.5.
(The lower end~0 is not natural, but recovers the conventional result;
the upper end~1.5 is where this order of chiral perturbation theory 
is less trustworthy.)
Since the typical range of fits leading to Eq.~(\ref{eq:latinput})
is $0.5<r<1.0$, we choose $r_0$ in this range and use Fig.~\ref{fig:xi}
to obtain
\begin{equation}
	\xi_f = 1.32 \pm 0.08.
	\label{eq:xifrange}
\end{equation}
With separation scale $\mu=1$~GeV, $\xi_f-1$ receives nearly equal
contributions from the low-energy constant (0.159) and the chiral
log~(0.165).%
\footnote{Loops with excited $B^{**}_q$ mesons are expected to
contribute significantly to~$\xi_f$~\cite{Falk:1993iu,Colangelo:1995ph},
but the ensuing $r$ dependence is well described by linear extrapolation,
so it is accurate to lump them into $(1-r)f_2(\mu)$.}

We have carried out a similar analysis for $\xi_B$ and also allowed for
a $\pm20\%$ range on~$g^2$.
(See the appendix for details.)
The chiral logs in $\xi_f$ and $\xi_B$ pull in opposite directions, so
the resulting $\xi=\xi_f\xi_B$ is insensitive to~$g^2$:
\begin{equation}
	\xi = 1.32 \pm 0.10,
	\label{eq:xirange}
\end{equation}
which is quite different from the range usually used in CKM~fits,
although it agrees with qualitative discussions of chiral
logs~\cite{Booth:1994hx,Sharpe:1996qp}, a direct analysis of
${\cal M}_s/{\cal M}_d$~\cite{Bernard:1998dg}, and chiral log fits to
preliminary unquenched calculations~\cite{Yamada:2001xp}.
The shift in central value from 1.15 to 1.32 can be thought of as a
correction to the quenched approximation: mature unquenched calculations
will certainly see the curvature required by the chiral log.

Because our result is so different than the conventional one,
let us stress the differences in methodology.
Usually $\xi$ is obtained via a linear chiral extrapolation, although 
chiral log fits have been tried in Ref.~\cite{Yamada:2001xp}.
We have relied completely on the functional form predicted by chiral 
perturbation theory. 
It is difficult to determine the coefficient of the chiral logs directly
from the lattice calculation.
We have circumvented this obstacle by using the $D^*$
width~\cite{Anastassov:2001cw}, which, with heavy-quark symmetry,
implies $g^2=0.35$.
The uncertainty in Eq.~(\ref{eq:xirange}) is larger than in many other
papers, mostly because we have assigned $\pm0.05$ instead of $\pm0.03$
uncertainty to the lattice calculations.
On the other hand, we have also not done a complete error analysis:
for example, we have neglected uncertainties from higher orders in the
chiral expansion.

One could easily reduce the theoretical uncertainty
in $B^0$-$\bar{B}^0$ mixing by carrying out lattice
calculations designed to determine the low-energy constants in
Eqs.~(\ref{eq:unquenchedfBs})--(\ref{eq:unquenchedBBd}).
If one takes closely-spaced values of the light quark mass, even if
close to the strange mass, one can compute the derivative~$d\xi/dr$.
If one is willing to take $g^2$ from experiment, these derivatives give
$f_2(\mu)$ and $B_2(\mu)$, and one can proceed to determine~$\xi$ for
physically light quark masses.
The same procedure could be applied to $f_{B_q}$ and $B_{B_q}$ although
now one must also compute $f_1(\mu)$ and $B_1(\mu)$, and also cope
with further low-energy constants in the $1/m_b$
corrections~\cite{Boyd:1994pa,Booth:1994rr}.
Chiral extrapolations with chiral logs may well change $f_{B_q}$ from
the estimates in Eqs.~(\ref{eq:fBs}) and~(\ref{eq:fBd}) in the same
way they changed $\xi_f$.

\begin{figure}[b]
	\centering
 	\includegraphics[width=0.8\textwidth]{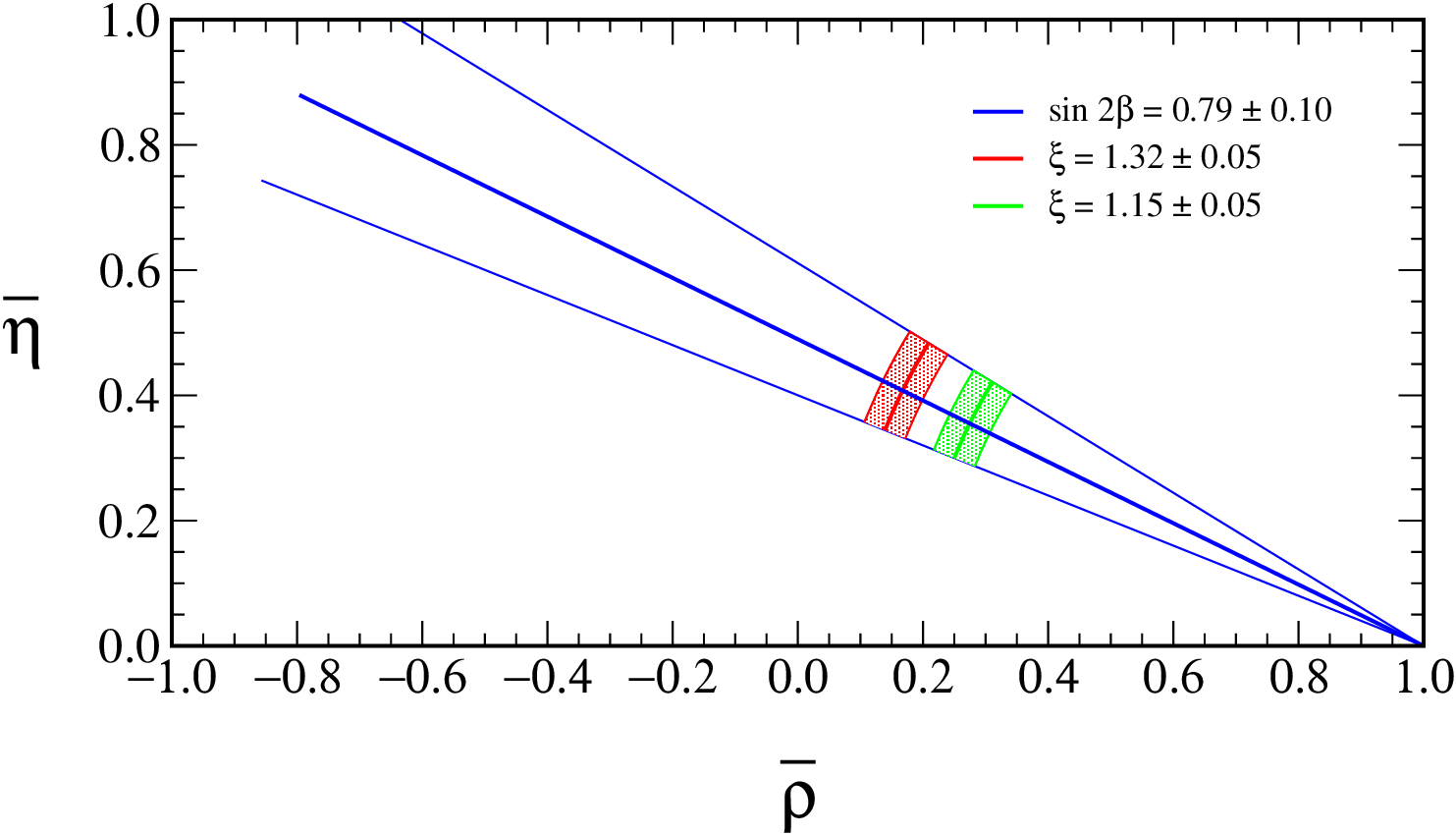}
	\caption[fig:ut]{Sketch of constraints on the apex of the
	unitarity triangle with $\sin 2\beta=0.79\pm0.10$,
	$\Delta m_s=20~\textrm{ps}^{-1}$ and
	$\xi=1.32\pm0.05$ or $1.15\pm0.05$.}
	\label{fig:ut}
\end{figure}

From a (lattice) purist's point of view it may be unsatisfactory to
take $g^2$ from experiment.
In the long run it will, however, be possible to solidify our knowledge
of $g^2$ (in the $B$ system) through lattice calculations and other
applications of chiral perturbation theory to $B$~physics.
To relate the very precise measurements to the CKM matrix, the
combination of phenomenology for $g^2$ and lattice calculation for
the low-energy constant is very satisfactory, especially since we find
that $\xi$ varies by less than 2\% when $g^2$ is varied by 20\%.
Fig.~\ref{fig:ut} shows how the combination of $\sin 2\beta$ and
$\Delta m_s/\Delta m_d$ work together to constrain the apex of the
unitarity triangle.
We take $\sin 2\beta=0.79\pm0.10$ from averaging
CDF~\cite{Affolder:1999gg}, BaBar~\cite{Aubert:2002rg} and
Belle~\cite{Abe:2002wn} measurements.
For illustration we take $\Delta m_s = 20~\textrm{ps}^{-1}$,
and compare $\xi=1.15\pm0.05$ (conventional wisdom) with
$\xi=1.32\pm0.05$ [Eq.~(\ref{eq:xirange}) with error halved].
With a larger value of $\xi$ the mixing side is longer,
scaling like $\xi/\sqrt{\Delta m_s}$.
By the same token, our result suggests that the Standard-Model
prediction for~$\Delta m_s$ (16--19~ps$^{-1}$ \cite{Ciuchini:2000de})
should be increased, perhaps by 25--35\%.

\vskip 2.0em
\noindent{\sl Acknowledgments:}
We would like to thank
Claude Bernard,
Gustavo Burdman,
Shoji Hashi\-moto,
Aida El-Khadra,
Zoltan Ligeti,
Vittorio Lubicz,
Ulrich Nierste,
and Norikazu Yamada
for discussions related to this work.

\newpage
\noindent
{\sl Appendix: Analysis including $\xi_B$ and varying $g^2$}
\vskip 1.0em

Let $\xi^2_B=B_{B_s}/B_{B_d}$, with linear chiral extrapolation
\begin{equation}
	\xi^2_B(r) - 1 = (1-r)S_B.
\end{equation}
Then, eliminating $B_2(\mu)$ in the same way as $f_2(\mu)$ in $\xi_f$,
\begin{equation}
	\xi^2_B(r) - 1 = (1 - r) \left\{S_B + m_{ss}^2 
	\frac{1-3g^2}{(4\pi f)^2} \left[ l_B(r_0) - l_B(r) \right] \right\},
	\label{eq:xiBfinal}
\end{equation}
where
\begin{equation}
	l_B(r) = \frac{1}{1-r}\left[
		\frac{2+r}{6} \ln\left(\frac{2+r}{3}\right) - 
		\frac{r}{2}   \ln(r) \right].
	\label{eq:chilogB}
\end{equation}
To evaluate the right-hand side, we take~\cite{Ryan:2001ej}
\begin{equation}
	S_B = 0.00 \pm 0.05.
\end{equation}
Then we find $\xi_B=0.998\pm0.025$.

In the main analysis, we have used $g^2=0.35$, which assumes that the
$B$-$B^*$-$\pi$ and $D$-$D^*$-$\pi$ are the same.
Repeating the analysis with $g^2=0.20$ and $0.50$ we find the results in
Table~\ref{tab:g2}.
Although the chiral extrapolation of $\xi_B$ is no longer completely
insignificant, and $\xi_f$ changes a little,
the result for~$\xi$ is very stable.
\begin{table}[h]
	\centering
	\caption[tab:g2]{Comparison of chiral extrapolations for $\xi_f$,
	$\xi_B$ and $\xi$ for three values of the $B$-$B^*$-$\pi$ coupling
	$g^2=0.20$, 0.35, 0.50.}
	\label{tab:g2}
	\vskip 2pt
	\begin{tabular}{cr@{$\pm$}lr@{$\pm$}lr@{$\pm$}l}
		\hline \hline
		  $g^2$ &	\multicolumn{2}{c}{$\xi_f$} &
					\multicolumn{2}{c}{$\xi_B$} &
					\multicolumn{2}{c}{$\xi$}  \\
		\hline
		  0.20  & 1.29&0.08 & 1.01&0.03  & 1.30&0.09 \\
		  0.35  & 1.32&0.08 & 1.00&0.02  & 1.32&0.09 \\
		  0.50  & 1.36&0.09 & 0.98&0.02  & 1.34&0.09 \\
		\hline \hline
	\end{tabular}
\end{table}

\end{document}